\documentclass[twocolumn,english,aps,pra,superscriptaddress,showpacs,tightenlines]{revtex4-1}
\usepackage{amsmath}
\usepackage{graphicx}
\usepackage{amssymb}
\usepackage{CJK}
\usepackage{color}
\usepackage{titlesec}
\usepackage{appendix}
\usepackage{mathrsfs}

\begin{document}

\begin{CJK*}{GBK}{song}

\title{Coherent Control of Spontaneous Emission for a giant driven $\Lambda $-type three-level atom}
\author{Ya \surname{Yang}}
\affiliation{Synergetic Innovation Center for Quantum Effects and Applications, Key Laboratory for Matter Microstructure and Function of Hunan Province, Key Laboratory of Low-Dimensional Quantum Structures and Quantum Control of Ministry of Education, School of Physics and Electronics, Hunan Normal University, Changsha 410081, China}
\affiliation{School of Physics and Chemistry, Hunan First Normal University, Changsha 410205, China}
\author{Ge \surname{Sun} }
\affiliation{Synergetic Innovation Center for Quantum Effects and Applications, Key Laboratory for Matter Microstructure and Function of Hunan Province, Key Laboratory of Low-Dimensional Quantum Structures and Quantum Control of Ministry of Education, School of Physics and Electronics, Hunan Normal University, Changsha 410081, China}
\author{Jing \surname{Li} }
\affiliation{Synergetic Innovation Center for Quantum Effects and Applications, Key Laboratory for Matter Microstructure and Function of Hunan Province, Key Laboratory of Low-Dimensional Quantum Structures and Quantum Control of Ministry of Education, School of Physics and Electronics, Hunan Normal University, Changsha 410081, China}
\author{Jing \surname{Lu} }
\affiliation{Synergetic Innovation Center for Quantum Effects and Applications, Key Laboratory for Matter Microstructure and Function of Hunan Province, Key Laboratory of Low-Dimensional Quantum Structures and Quantum Control of Ministry of Education, School of Physics and Electronics, Hunan Normal University, Changsha 410081, China}
\author{Lan \surname{Zhou}}
\affiliation{Synergetic Innovation Center for Quantum Effects and Applications, Key Laboratory for Matter Microstructure and Function of Hunan Province, Key Laboratory of Low-Dimensional Quantum Structures and Quantum Control of Ministry of Education, School of Physics and Electronics, Hunan Normal University, Changsha 410081, China}
\begin{abstract}
Quantum optics with giant atoms provides a new approach for implementing optical memory devices at the atomic scale. Here, we theoretically study the relaxation
dynamics of a single driven three-level atom interacting with a
one-dimensional waveguide, via two coupling points.
Under certain conditions, after the long-time dynamics, we found that the population of giant atom can either maintain stable values or exhibit regular periodic oscillation behavior, while photons can be trapped in the region of giant atoms.
This phenomenon is not achievable using a two-level atom with two legs.
It is worth noting that the atomic excitation probability of a stable bound state is a constant value, which is determined by the size of the atom.
Crucially, the size of the atom (the distance between the two coupling points) is much larger than the wavelength of the light field, which is a necessary condition for the existence of oscillating bound states.

\end{abstract}

%42.50.Pq, 42.50.Ex, 03.67.Lx, 78.67.-n
\maketitle

\end{CJK*}\narrowtext
\renewcommand\thesection{\arabic{section}}

\section{Introduction}

In traditional quantum optics, the wavelength of the light is typically much larger than the atomic scale. The atoms in nature are very small $\left(  r\approx10^{-10}\mathrm{m}\right)  $, and we focus on the interaction between these atoms and the electromagnetic waves of optical wavelengths $\left(  \lambda\approx10^{-6}-10^{-7}\mathrm{m}\right)  $\cite{D. Leibfried03,R. Miller05,A. F. Kockum14}. So that atoms are usually considered point-like quantum emitters, and the atom-field coupling is local within the dipole approximation\cite{S. Haroche06,A. F. Kockum14}. For an excited atom emitting in a featureless continuum of electromagnetic modes in the vacuum state, the resulting process of spontaneous emission is well described by the Weisskopf-Wigner theory\cite{V. F. Weisskopf30}. According to this theory, the excited atomic state undergoes an exponential decay to the ground state, accompanied by the irreversible emission of a single photon\cite{V. F. Weisskopf30,S. Longhi20}.
However, following significant technological advances in superconducting circuits\cite{J. Q. You11,X. Gu17,A. F. Kockum19-1,P. Krantz19}, "giant" artificial atoms\cite{A. F. Kockum21,S. Longhi20,W. Zhao20,H. Yu21} such as transmon qubits (TQs)\cite{J. Koch07,R. Barends13} have been designed. These TQs couple to surface acoustic waves (SAWs) via multiple points spaced wavelength distances apart\cite{M. V. Gustafsson14,T. Aref16,A. Noguchi17,R. Manenti17,A. N. Bolgar17,B. A. Moores18,K. J. Satzinger18,G. Andersson19,L. R. Sletten19,A. Bienfait19}, rather than behaving as point-like emitters as assumed in the traditional Weisskopf-Wigner theory.

Giant atoms with dimensions exceeding the wavelength of photons exhibit a decay behavior that departs from the conventional exponential decay model\cite{G. Andersson19,S. Longhi20}. Photons in the ground state are capable of reversible transitions back to the excited state, indicating a departure from the standard spontaneous emission dynamics observed in smaller atomic systems. The propagation time of light between coupled points cannot be ignored, implying that the evolution of the system does not follow a Markovian process. By analyzing the delayed evolution between coupling points, we gain a deeper understanding of this non-Markovian effect.

Research on non-Markovian processes in various quantum systems has gained momentum with the advancement of quantum information technology. Studies have revealed that non-Markovian retarded effects induced by significant spatial separations. These effects have been observed in scenarios like a single atom and the waveguide end (i.e., single small atom in front of a mirror)\cite{J. Eschner01,U. Dorner02,T. Tufarelli13,T. Tufarelli14,P. O. Guimond17,G. Calajo19}, distinct coupling ports of gaint atoms\cite{S. Longhi20,L. Guo20,L. Du21-1,W. Zhao20,X. Zhang23}, and interactions among distant atoms\cite{P. W. Milonni74,H. Zheng110,C. Gonzalez-Ballestero13,M. Laakso14,K. Sinha20,S. Longhi20,A. Carmele20,L. Du21,P. O. Guimond16,T. Ramos16,F. Dinc19,C. Gonzalez-Ballestero14}.

In this paper, we explore the relaxation dynamics of a giant driven $\Lambda $-type three-level atom interacting with a one-dimensional (1D) open waveguide through two coupling points. Our key finding is that when the distance between the two coupling points of a giant atom exceeds the wavelength of the field, the time for photon propagation between them becomes significant. Under specific conditions, this distance results in the emergence of a bound state and a quasi-bound state in the system. The interference between these states alters the decay of spontaneous emission from exponential to polynomial, displaying oscillatory behavior known as non-Markov effects. With further distance increase, particularly around 100 wavelengths, spontaneous emission exhibits prolonged and sustained periodic oscillations.
The presence of two coupling points allows for the creation of persistently oscillating bound states, a phenomenon that, to our knowledge, is unique to three-level atoms. We anticipate that this phenomenon could be utilized for storing and manipulating quantum information in larger Hilbert spaces and could be seen as a simplified implementation of cavity QED, with the atom effectively forming its own cavity. As a result, such a system may be relevant for a single atom,
 optical memory device.

\section{\label{Sec:2} THE HAMILTONIAN AND EQUATIONS OF MOTION}

As schematically shown in Fig. 1, we consider a single giant $\Lambda$-type
three-level atom interacting with an open one-dimensional (1D) waveguide at
two separated points with the separation distance comparable to the wavelength
of the waveguide field. The Hamiltonian for the combined atom-waveguide system
as $H=H_{w}+H_{a}+H_{int}$, where $H_{w}$ and $H_{a}$ represent the free
Hamiltonian of the waveguide and the atom, respectively. $H_{int}$ denotes the
coupling between the waveguide and the atom. Within the rotating-wave
approximation, we write the Hamiltonian as (we set $\hbar=1$ throughout)
\begin{subequations}
\label{1-1}%
\begin{align}
H_{w}  &  =\int_{-\infty}^{+\infty}dk\omega_{k}a_{k}^{\dag}a_{k},\\
H_{a}  &  =\omega_{e}\sigma_{ee}+\omega_{s}\sigma_{ss}+\omega_{g}\sigma
_{gg}+\Omega e^{\mathrm{i}\omega_{l}t}\sigma_{se}+H.c.,\\
H_{int}  &  =\int_{-\infty}^{+\infty}dkg_{0}\left(  \left(  1+e^{\mathrm{i}%
kd}\right)  a_{k}\sigma_{eg}+H.c.\right)  \sqrt{\left\vert k\right\vert }.
\end{align}
The dispersion relationship of the photon with the wave vector $k$ in the
waveguide is approximately linear around the transition frequency $\omega
_{eg}\equiv\omega_{e}-\omega_{g}$ as $\omega_{k}=\omega_{eg}+v\left\vert
k-k_{0}\right\vert $, where $v$ is the group velocity of the
photons (setting $v=1$). The photon annihilation operator $a_{k}$ satisfy
$\left[  a_{k},a_{k^{\prime}}^{\dag}\right]  =\delta\left(  k-k^{\prime
}\right)  $. And the $\Lambda$-type atom contains the ground state $\left\vert
g\right\rangle $, excited state $\left\vert e\right\rangle $ and metastable
state $\left\vert s\right\rangle $ with eigenfrequencies $\omega_{g},$
$\omega_{e}$ and $\omega_{s}$, respectively. We defined the atomic operator
$\sigma_{ij}=\left\vert i\right\rangle \left\langle j\right\vert ,\left(
i,j=g,s,e\right)  .$ In addition, we contemplate that the excited state
$\left\vert e\right\rangle $ of the $\Lambda$-type atom (Fig. 1) is coupled to
metastable level $\left\vert s\right\rangle $ by a classical laser beam with
frequency $\Omega$. And the atomic transition $\left\vert g\right\rangle
\leftrightarrow\left\vert e\right\rangle $ coulped with the waveguide at two
coupling sites, which are set as $x_{1}=0$ and $x_{2}=d\equiv v\tau$,
respectively. And $\tau$ is the travel time for photons between the two
coupling points which characterized the size of giant atoms.
\begin{figure}[t!]
\includegraphics[width=8cm]{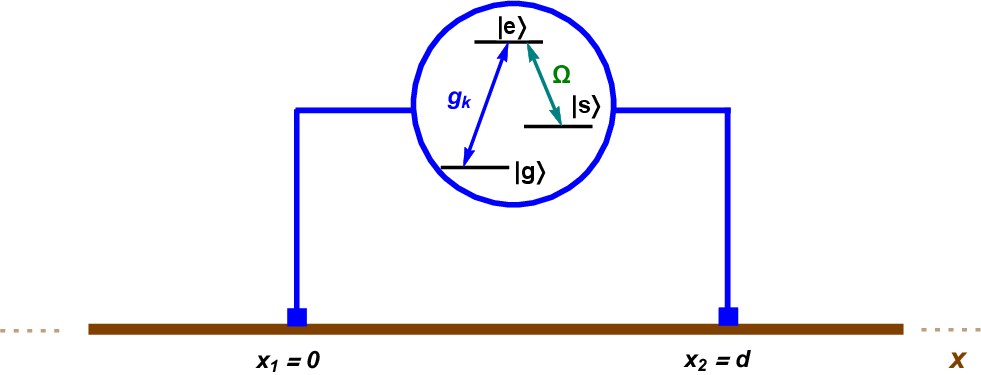}\caption{(Color online) Schematic of
model architecture. A $\Lambda$-type three-level atom interacts with a 1D
waveguide via two legs, and the distance between the two legs is $d$. The
transition $\left\vert e\right\rangle \leftrightarrow\left\vert g\right\rangle
$ is coupled to the waveguide with the coupling coefficient $g_{k}$. A
classical driving with an angular frequency $\omega_{l}$ and a Rabi frequency
$\Omega$ is applied to the transition $\left\vert e\right\rangle
\leftrightarrow\left\vert s\right\rangle $. }%
\end{figure}The last two items of $H_{a}$ depicts the interaction between the
three-level atom and external driving field, the transition $\left\vert
e\right\rangle \leftrightarrow\left\vert s\right\rangle $ is driven by the
classical field with driving strength $\Omega$\ and driving frequency
$\omega_{l}$. Making unitary transformation $U=\exp\left[  -\mathrm{i}%
\frac{\omega_{l}t}{2}\left(  \sigma_{ee}-\sigma_{ss}+\int_{-\infty}^{+\infty
}dka_{k}^{\dag}a_{k}\right)  \right]  $, we obtain
\end{subequations}
\begin{align}
H  &  =\alpha_{e}\sigma_{ee}+\alpha_{s}\sigma_{ss}+\omega_{g}\sigma_{gg}%
+\int_{-\infty}^{+\infty}dk\tilde{\omega}_{k}a_{k}^{\dag}a_{k}+\Omega
\sigma_{es}\nonumber\\
&  +\int_{-\infty}^{+\infty}dkg_{0}\sqrt{\left\vert k\right\vert }\left(
1+e^{\mathrm{i}kd}\right)  \sigma_{eg}a_{k}+H.c. \label{1-2}%
\end{align}
Here $\alpha_{e}=\omega_{e}-\omega_{l}/2,\alpha_{s}=\omega_{s}+\omega_{l}/2$
and $\tilde{\omega}_{k}=\omega_{k}-\omega_{l}/2.$ The total number of atomic
and field excitations is conserved. The system's state at time $t$ can be
expanded in the single-excitation subspace as%
\begin{equation}
\left\vert \psi\left(  t\right)  \right\rangle =C_{e}\left(  t\right)
\left\vert e,0\right\rangle +C_{s}\left(  t\right)  \left\vert
s,0\right\rangle +\int dk\beta
_{k}\left(  t\right)  \left\vert g,1_{k}%
\right\rangle , \label{1-3}%
\end{equation}
where $C_{e}\left(  t\right)  $, $C_{s}\left(  t\right)  $, and $\beta
_{k}\left(  t\right)  $ denote the atomic probability amplitudes in the states
$\left\vert e\right\rangle ,$ $\left\vert s\right\rangle $ and $\left\vert
g\right\rangle $, respectively. $\left\vert 1_{k}\right\rangle $ stand for
here includes a single photon with frequency $\omega_{k}$ in the waveguide.
From the Schr\"{o}inger equation, the derivation details of Eq. (2) can be
found in Appendix A, we derive the equation of motion for the probability
amplitude of the giant atom being excited state
\begin{align}
\frac{d}{dt}C_{e}\left(  t\right)   &  =\left(  -\mathrm{i}\alpha_{e}%
-\gamma\right)  C_{e}\left(  t\right)  -\gamma C_{e}\left(  t-\tau\right)
e^{-\mathrm{i}\phi}\Theta\left(  t-\tau\right) \nonumber\\
&  -\Omega^{2}\int_{0}^{t}dt^{\prime}C_{e}\left(  t^{\prime}\right)
e^{\mathrm{i}\alpha_{s}\left(  t^{\prime}-t\right)  }. \label{1-4}%
\end{align}
Here, the relaxation rate at single coupling point $\gamma\equiv4\pi
\omega_{eg}g_{0}^{2}$ can be approximated as a constant over the relevant
frequency range in the spirit of Weisskopf-Wigner theory. The first term on
the right-hand side of Eq. (\ref{1-3}) describes the free evolution of the
excited level and the non time delayed when the photon released from the mth
coupling point and then reabsorbed at the same point. The second term depicts
the relaxation processes due to the time delay between different coupling
points. The last term indicates that the external driving field coherently
drives the atom, which causes the transition between the levels $\left\vert
e\right\rangle $ and $\left\vert s\right\rangle $. In the two-level-atom
scenario $\left(  \Omega=0\right)  $, we recover the results given in Ref.
[5]. We assume that the giant atom (waveguide) is initially in the excited
$\left\vert e\right\rangle $ (vacuum $\left\vert 0\right\rangle $) state,
i.e., $C_{e}\left(  0\right)  =1$. Then, we obtain the solution of
$\beta\left(  t\right)  $ by a Laplace transformation:
\begin{equation}
C_{e}\left(  t\right)  =\sum\limits_{p}\frac{e^{s_{p}t}}{1-\gamma\tau
e^{-\mathrm{i}\phi}e^{-s_{p}\tau}-\Omega^{2}/\left(  s_{p}+\mathrm{i}%
\alpha_{s}\right)  ^{2}}, \label{1-5}%
\end{equation}
and the complex frequency parameters $s_{p}$ are given by
\begin{equation}
s_{p}+\mathrm{i}\alpha_{e}+\gamma+\gamma e^{-\mathrm{i}\phi}e^{-s_{p}\tau
}+\frac{\Omega^{2}}{s_{p}+\mathrm{i}\alpha_{s}}=0. \label{1-6}%
\end{equation}
Furthermore, the photonic population in the waveguide $\varphi\left(
x,t\right)  \equiv\frac{1}{\sqrt{2\pi}}\int_{-\infty}^{+\infty}dke^{\mathrm{i}%
kx}\beta_{k}\left(  t\right)  $ in the real space is
\begin{align}
\varphi\left(  x,t\right)    & =-\mathrm{i}\sqrt{\frac{\gamma}{2}}C_{e}\left(
t-\frac{\left\vert x-d\right\vert }{v}\right)  e^{-\mathrm{i}\phi
\frac{\left\vert x-d\right\vert }{d}}\Theta\left(  t-\frac{\left\vert
x-d\right\vert }{v}\right)  \nonumber\\
& -\mathrm{i}\sqrt{\frac{\gamma}{2}}C_{e}\left(  t-\frac{\left\vert
x\right\vert }{v}\right)  e^{-\mathrm{i}\phi\frac{\left\vert x\right\vert }%
{d}}\Theta\left(  t-\frac{\left\vert x\right\vert }{v}\right)  ,\label{1-7}%
\end{align}
with $\phi=\left(  \omega_{g}-\omega_{l}/2\right)  \tau$ and the time
evolution of the atomic probability amplitude in the metastable state
$\left\vert s\right\rangle $ is
\begin{equation}
C_{s}\left(  t\right)  =-\mathrm{i}\Omega\int_{0}^{t}dt^{\prime}C_{e}\left(
t^{\prime}\right)  e^{\mathrm{i}\alpha_{s}\left(  t^{\prime}-t\right)  }.
\label{1-8}%
\end{equation}
In general, for finite time delay $\tau>0$, the nonlinear Eq. (\ref{1-5}) has
multiple solutions and has not simple closed form.

\section{\label{Sec:3} DISCUSSION OF BOUND-STATE}
\begin{figure*}[t!]
  \centering
  \includegraphics[width=16cm]{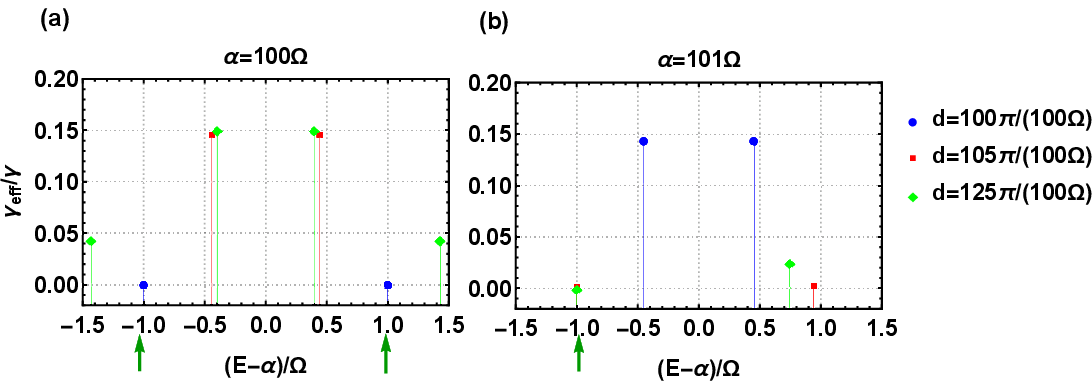}\\
  \caption{The figure shows the poles by solving Eq. (\ref{1-6}). The vertical axis represents the negative real part of the pole, which signifies the decay rate of the giant $\Lambda$-type three-level atom with two coupling points separated by a distance $d=\frac{\pi}{\Omega}$ [blue-point], $d=\frac{105\pi}{100\Omega}$ [red-point], $d=\frac{125\pi}{100\Omega}$ [green-point]. And the horizontal axis represents the imaginary part of the pole offset by $\alpha$, indicating energy shift. Here, we assumed that $\phi=0$, $\Delta=0$, $\alpha_{e}=\alpha_{s}=\alpha$ and $\gamma=\Omega$. All the parameters are in units of $\Omega$. The parameters are set as (a) $\alpha=100\Omega$ and (b) $\alpha=101\Omega$.}
  \label{f-2}
\end{figure*}
Physically, the poles $s_{p}$ have a negative real part, which represents the
decay rate. In some particular situations, the so-called dark state or bound
state exists which does not decay despite the dissipative environment. The
corresponding poles $s_{p}$ are pure imaginary. We set the purely imaginary
solution $s_{p}\equiv-\mathrm{i}\Omega_{p}$ with (see Appendix B), and we obtain the frequencies of the two dark modes: $\Omega_{p}=\omega_{\pm
},\omega_{\pm}=\left(  \Delta\pm\Omega_{eff}\right)  /2,$ with the detuning
$\Delta\equiv$ $\alpha_{e}-\alpha_{s},$ and the effect frequency $\Omega
_{eff}\equiv\sqrt{\left(  \alpha_{e}-\alpha_{s}\right)  ^{2}+4\Omega^{2}}.$ We
obtain the corresponding bound-state conditions satisfy
\begin{subequations}
\label{3-2}%
\begin{align}
\Omega_{p_{1}} &  =\omega_{+}=\frac{\left(  2p_{1}+1\right)  \pi+\phi}{\tau
},\\
\Omega_{p_{2}} &  =\omega_{-}=\frac{\left(  2p_{2}+1\right)  \pi+\phi}{\tau},
\end{align}
\end{subequations}
where $p_{1}$ and $p_{2}\in Z.$\ It is possible to find two integers $p_{1}$
and $p_{2}$ satisfying Eq. (\ref{1-6}) simultaneously. And also possible to
find only one integer $p_{1}$ or $p_{2}$.

Next, we will find the conditions for two solutions to coexist. For the sake
of simplicity, let's assume $\phi=0,\Delta=0,\alpha_{e}=\alpha_{s}=q\Omega$,
which is not a general discussion. We obtain the following condition for the
finite time delay
\begin{equation}
\tau=\frac{\left(  2p_{1}+1\right)  \pi}{\left(  q+1\right)  \Omega}%
=\frac{\left(  2p_{2}+1\right)  \pi}{\left(  q-1\right)  \Omega}. \label{3-3}%
\end{equation}
If $q$ is a positive integer, it must be an even number, and then the integers
$p_{1}$ and $p_{2}$ are obtained
\begin{subequations}
\label{3-4}%
\begin{align}
p_{1}  &  =\frac{q}{2}+\left(  q+1\right)  n,\\
p_{2}  &  =\frac{q}{2}-1+\left(  q-1\right)  n,
\end{align}
and the distance between two legs is%
\end{subequations}
\begin{equation}
d=\left(  2n+1\right)  \frac{\pi}{\Omega}, \label{3-5}%
\end{equation}
where\ $n\in N.$ This indicates that the size of atom being much larger than
the wavelength of the photon is necessary for the existence of two bound
states. If $q$ is a decimal, there will also be certain constraints on $q$,
and a detailed discussion can be found in Appendix C.

By solving Eq. (\ref{1-6}), with $\phi=0$ and $\Delta=0$, we depict the decay rate and energy shift as functions of the giant atom's size in Fig. \ref{f-2}(a)-(b). In Fig. \ref{f-2}(a) with $\alpha=100\Omega$, it is observed that for $d=\frac{\pi}{\Omega}$ [blue-point], two distinct solutions exhibit zero decay rates, corresponding precisely to the coexistence of energies energies $\omega_{+}$ and $\omega_{-}$. But under the same size $d=\frac{\pi}{\Omega}$, in Fig. \ref{f-2}(b), there is no dark mode.
Besides, in Fig. \ref{f-2}(b) with $\alpha=101\Omega$, at distances $d=\frac{105\pi}{100\Omega}$ [red-point], and $d=\frac{125\pi}{100\Omega}$ [green-point], only one dark state solution with a zero decay rate is observed.

\subsection{STATIC BOUND STATES}
\begin{figure*}[tphb]
\includegraphics[width=16cm]{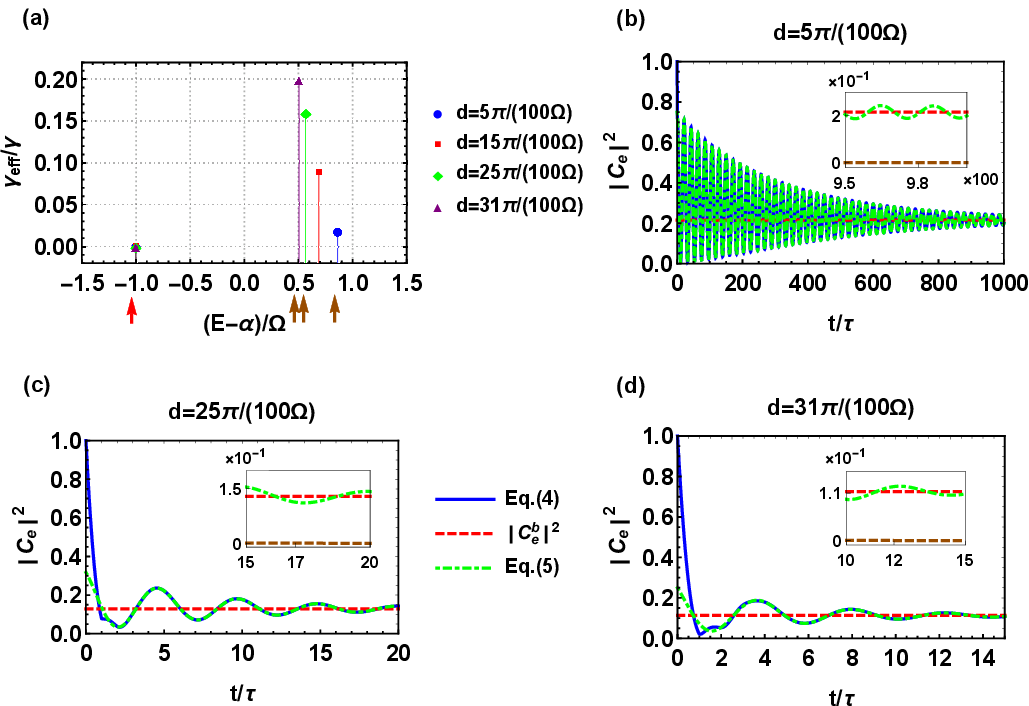}\caption{(Color online) (a) displays the poles obtained by solving Eq. (\ref{1-6}) with $\phi=0$ , $\Delta=0$ and $\alpha=101\Omega$. The different
color points show the poles for the different distances. (b)-(d) depict the probability $\left\vert C_{e}\left(  t\right)  \right\vert ^{2}$ for the giant atom in the excited state as a function of scaled time $t/\tau$ under distances $d=\frac{5\pi}{100\Omega}$ , $d=\frac{25\pi}{100\Omega}$, $d=\frac{31\pi}{100\Omega}$, respectively. The blue solid, red dashed and green dot-dashed lines in the planes (b)-(d) represent the numerical solution for the Eq. (\ref{1-4}) and the dark mode solution in Eq. (\ref{1-13}), as well sa the analytical solution for the Eq. (\ref{1-5}), respectively. The insets provide a detailed view of the probability amplitudes over an extended period. The red dashed lines illustrate the dark mode characterized by pure imaginary poles, while the brown dashed lines represent the decaying quasi-bound state mode with complex poles. Within the insets, the green dot-dashed lines showcase the synergistic impact of the dark mode and the propagation mode.
}
\label{fig2}
\end{figure*}

We consider the static bound state satisfying case: the integer $p_{1}$
exists, $p_{2}$ does not exist, or $p_{2}$ exists and $p_{1}$ does not exist,
which means that the bound state in the waveguide has only one frequency. Let us first discuss the situation of $\phi=0$ , $\Delta=0$ and $\alpha=101\Omega$. In this case,
 the unique pure imaginary solution is $s_{p}=-\mathrm{i}\omega_{-
}$. Substituting the solution into Eq. (\ref{1-5}), the dynamics of the atomic excitation
probability amplitude for the bound state reads
\begin{equation}
C_{e}^{b}\left(  t\right)  =\frac{1}{\gamma\tau+2}e^{-\mathrm{i}\left(
\alpha-\Omega\right)  t},
\label{1-13}
\end{equation}
and the bound state atomic probability amplitude population on the state
$\left\vert s\right\rangle $ is%
\begin{equation}
C_{s}^{b}\left(  t\right)  =-\frac{1}{\gamma\tau+2}e^{-\mathrm{i}\alpha
t}\left(  e^{\mathrm{i}\Omega t}-1\right)  .
\end{equation}
then the explicit expression for the photonic excitation probability density function $p\left(  x,t\right)
=\left\vert \varphi\left(  x,t\right)  \right\vert ^{2}$ in the site $x$ for the bound state in the long-time limit
\begin{equation}
p\left(  x,t\right)  =\frac{\gamma}{2}\frac{1}{\left(  \gamma\tau+2\right)
^{2}}\left\vert 1+e^{\mathrm{i}\left(  \alpha-\Omega\right)  \frac{\left\vert
x-d\right\vert -\left\vert x\right\vert }{v}}\right\vert ^{2},
\end{equation}
When the position $x$ is outside the interval $\left[  0,d\right]  $, because
the value of $\left(  \left\vert x-d\right\vert -\left\vert x\right\vert
\right)  /\left(  v\tau\right)  $ is fixed, either $1$ for $x\leq0$ or
$-1$\ for $x\geq d$. This leads to the establishment of $p\left(  x,t\right)
=0,x\in(-\infty,0]$
or $x\in\lbrack d,+\infty)$. That to say, outside the giant atom, the field
probability distribution $p\left(  x,t\right)  $ is zero, there is an
associated bound field state in the waveguide for a given dark state of the atom.

However, the poles $s_{p}$ that satisfies Eq. (\ref{1-6}) are not only uniquely pure imaginary, but also comprise a set of complex solutions shown in Fig. \ref{f-2}(b). We substitute the pure imaginary solution $s_{p}=-\mathrm{i}\omega_{-
}$
and the complex
solution $s_{p}=a+\mathrm{i}b$ into Eq. (\ref{1-5}), the dynamics of the atomic
excitation probability amplitude reads
\begin{subequations}
\label{3-10}%
\begin{align}
C_{e}\left(  t\right)    & =C_{e}^{b}\left(  t\right)  +C_{e}^{p}\left(
t\right)  ,\label{3-10a}\\
C_{e}^{p}\left(  t\right)    & =\frac{e^{at}e^{\mathrm{i}bt}}{1-\gamma\tau
e^{-a\tau}e^{-\mathrm{i}b\tau}-\Omega^{2}/\left[  a+\mathrm{i}\left(
b+\alpha\right)  \right]  ^{2}}.
\end{align}
\end{subequations}

As depicted in Fig. \ref{fig2}(a), we assume the emitter's frequency is $\Delta =0$ and $\alpha = 101\Omega$ (refer to Fig. \ref{f-2}(b)). Notably, a dark mode with an effective decay rate of
$\gamma_{eff}=0$ persists regardless of the atomic size, as indicated by the red arrow on the left side of Fig. \ref{fig2}(a). Besides, a series of brown arrows on the right side of Fig. \ref{fig2}(a) depict a complex solution at different distances $d$. The graph shows that as the distance increases, the decay rate $\gamma_{eff}$ also increases, indicating faster attainment of a steady state for the atoms. Through prolonged evolution, the attenuation term of atomic spontaneous emission gradually decreases to zero, as indicated by the brown dashed line in the inset of Figs. \ref{fig2}(b)-(d).
When the number of molecules in $d$ is odd, the coupling strength of a dark mode nearly disappears. Consequently, most of the energy remains trapped within the transmitter, with only a small amount escaping into the waveguide. For simplicity, we use Bound State (BS) to refer to the dark mode and quasi-bound state qBS for the complex solution.

We compare the analytical solutions from Eq. (\ref{3-10a}) with numerical simulations based on Eq. (\ref{1-4}) by plotting the atomic excitation probability $\left\vert C_{e}\left(  t\right)  \right\vert ^{2}$ for $\alpha = 101\Omega$ against the scaled time $t/\tau$ for various distances between the two coupling points in Figs. \ref{fig2}(b)-(d).
The blue solid, green dot-dashed, and red dashed lines correspond to the numerical, analytical, and purely imaginary results, respectively. It is observed that the model incorporating complex solution better matches the numerical result than the purely imaginary model. In Figs. \ref{fig2}(b)-(d), the atomic excitation probability $\left\vert C_{e}\left(  t\right)  \right\vert ^{2}$ in all three cases eventually stabilizes at a non-zero value after a long period, indicating the capture of the photon and the formation of a bound state, attributed to the photon transfer at the two coupling points.

It is easy to see that the early oscillations of the atomic population $\left\vert C_{e}\left(  t\right)  \right\vert ^{2}$ is the combined effect of the down dressed state $\omega_{-}$ (the purely imaginary result) and the up dressed state $\left(\omega_{+}-\gamma/2\right)$ (the negative imaginary part $b$ of the complex result). As time approaches infinity, the negative real part $a$ of the complex pole, representing the relaxation rate, leads to the contribution of the complex pole approaching zero. This interference results in oscillatory behavior in the spontaneous emission, as shown by the blue solid and green dot-dashed lines in Figs. \ref{fig2}(a) and \ref{fig2}(b). As a result of interference between the BS and qBS , the spontaneous
emission displays oscillatory behavior as depicted in blue solid and green dot-dashed line in
Figs. \ref{fig2}(b)-(d). In the long time limit, only the first term in Eq. (\ref{3-10}) remains dominant. The atomic steady-state population on the exited state $\left\vert
e\right\rangle$ can be then written as
\begin{equation}
\left\vert C_{e}\left(  t\right)  \right\vert ^{2}=\frac{1}{\left(  \gamma
\tau+2\right)  ^{2}},\label{3-11}
\end{equation}
Clearly, the steady-state atomic population is strongly dependent on the size of the giant atom. This means that the static bound state with different probability can be got through tuning
the distance of the two coupling points.

\subsection{OSCILLATING BOUND STATES}
\begin{figure*}[t!]
\includegraphics[width=16cm]{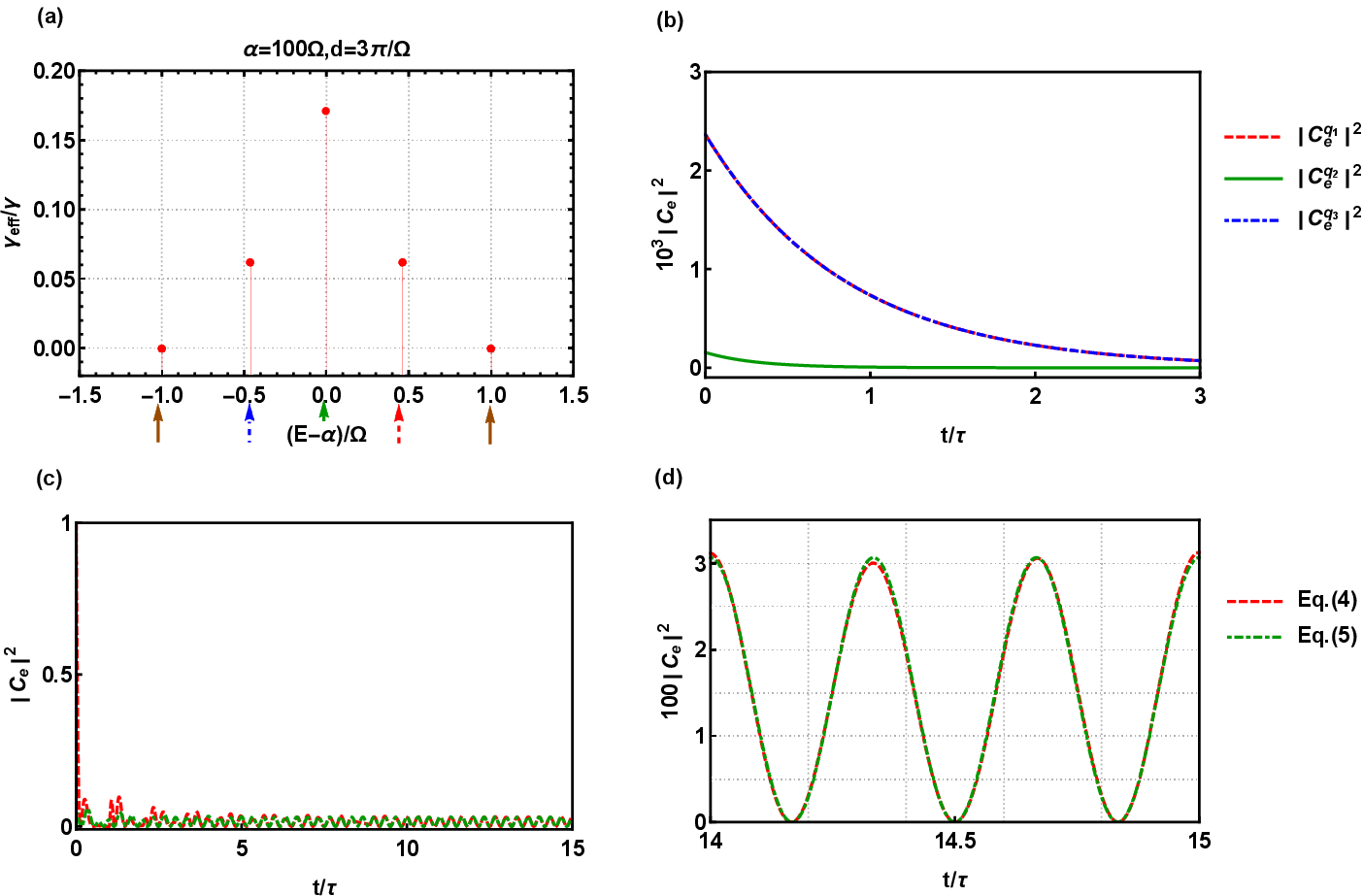}\caption{(Color online) (a) displays the poles obtained by solving Eq. (\ref{1-6}) with $\phi=0$, $\Delta=0$, $\alpha=100\Omega$ and $d=\frac{3\pi}{\Omega}$. (b) depicts how three complex poles in the (a) plane contribute to the probability of atomic excitation as a function of the scaled time $t/\tau$. (c)-(d) Evolution of atomic excitation probability based on numerical calculation (red dashed line) in Eq. (\ref{1-4}) and the analytical calculation (green dot-dashed line) in Eq. (\ref{1-5}). The parameters are the same as those in plane (a).}
\label{fig3}
\end{figure*}
In Fig. \ref{f-2}(a), the system demonstrates the coexistence of two bound states with frequencies $\omega_{+}$ and $\omega_{-}$ when satisfying Eq. (\ref{3-5}). By substituting $s_{p}=-\mathrm{i}\omega_{+}$ and $s_{p}=-\mathrm{i}\omega_{-}$ into Eq. (\ref{1-5}), the dynamics of the atomic excitation probability amplitude for these two bound states can be described as:
\begin{figure*}[t!]
\includegraphics[width=16cm]{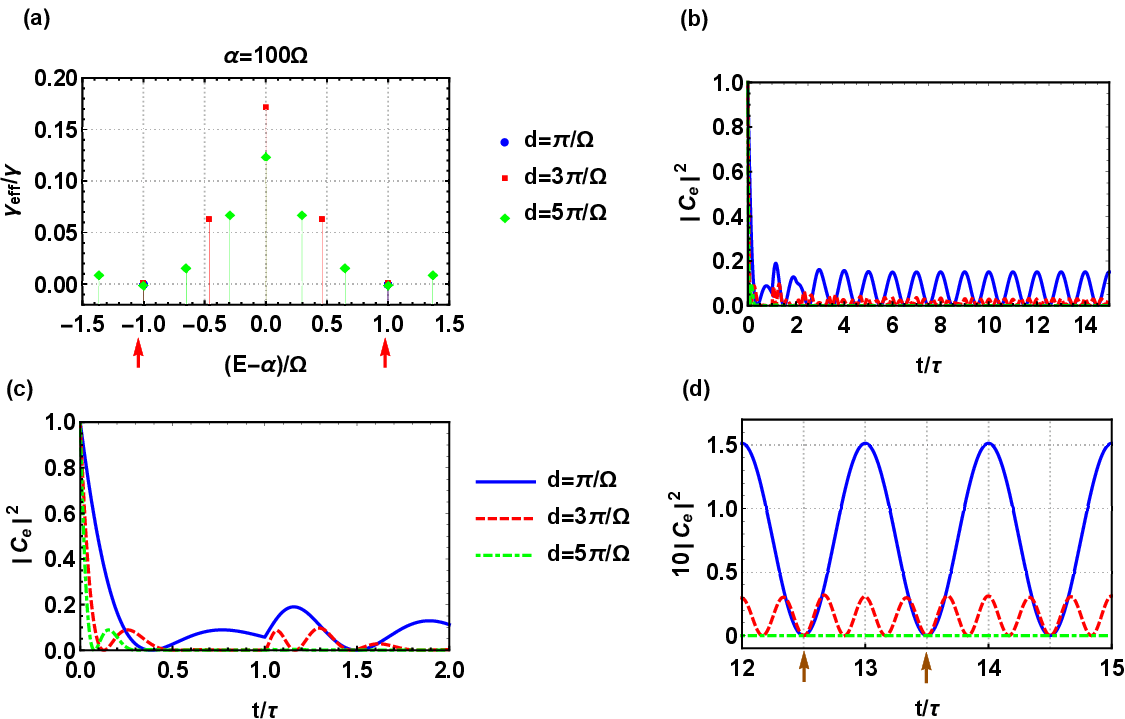}\caption{(Color online) (a) displays the poles obtained by solving Eq. (\ref{1-6}) with $\phi=0$ , $\Delta=0$ and $\alpha=100\Omega$. The different
color points show the poles for the different distances. (b)-(d) Evolution of the excited-state probability of a giant emitter for three different distances between two coulping points. }
\label{fig4}
\end{figure*}
\begin{equation}
C_{e}^{b}\left(  t\right)  =\frac{2}{\gamma\tau+2}e^{-\mathrm{i}\alpha t}%
\cos\left(  \Omega t\right)  ,\label{3-12}%
\end{equation}
and the field density in the long-time limit is
\begin{widetext}
\begin{equation}
p\left(  x,+\infty\right)  =\frac{\gamma}{2}\frac{4}{\left(  \gamma
\tau+2\right)  ^{2}}\left\vert
\begin{array}
[c]{c}%
\cos\left[  \Omega\left(  t-\frac{\left\vert x\right\vert }{v}\right)
\right]  \left[  1+e^{\mathrm{i}\alpha\frac{\left\vert x-d\right\vert
-\left\vert x\right\vert }{v}}\cos\left(  \Omega\frac{\left\vert
x-d\right\vert -\left\vert x\right\vert }{v}\right)  \right]  \\
+e^{\mathrm{i}\alpha\frac{\left\vert x-d\right\vert -\left\vert x\right\vert
}{v}}\sin\left(  \Omega\left(  t-\frac{\left\vert x\right\vert }{v}\right)
\right)  \sin\left(  \Omega\frac{\left\vert x-d\right\vert -\left\vert
x\right\vert }{v}\right)
\end{array}
\right\vert ^{2}.
\end{equation}
Similarly, when calculating the photon probability within the interval $\left(  d,+\infty\right)  $ or $\left(  -\infty,0\right)  $, the expression $\left(  \left\vert x-d\right\vert -\left\vert x\right\vert \right)
/v$ equals $\mp\tau $. This leads to $\mp\tau \alpha=2k\pi, k\in N$ and $\mp\tau\Omega=\left(2n+1\right)\pi, n\in N$. Therefore, outside the giant atoms, the field's probability distribution $p\left(x, t\right)$ is zero,
indicating that photons are localized between the two coupling points of the giant atoms.
\end{widetext}

As depicted in Fig. \ref{f-2}(a), it adheres to Eq (\ref{1-6}). The poles $s_{p}$ encompass a range of complex solutions beyond the two pure imaginary solutions.
In Fig. \ref{fig3}(a), we plot the poles for the Eq. (\ref{1-6}) with $\phi=0$, $\Delta=0$, $\alpha=100\Omega$ and $d=\frac{3\pi}{\Omega}$. At this juncture, the image reveals the existence of two symmetric dark states (two brown arrows) and three symmetric quasi-bound states. To analyze the influences of the three quasi-bound states, we plotted Fig. \ref{fig3}(b) and employed the inverse Laplace transform of the three complex poles to derive their contribution functions $\left\vert C_{e}^{q_{i}}\left(  t\right)  \right\vert ^{2}$ ($i=1,2,3$) over time scaled time $t/\tau$. The amplitudes of the three quasi-bound states are extremely small, each on the order of $10^{-3}$, almost negligible. Moreover, these quasi-bound states rapidly decay to zero. Hence, within a brief time frame, the evolution of the probability of atomic excited state transforms into an oscillation driven by two bound states. Figs. \ref{fig3}(c)-(d) shows the excitation dynamics of the giant atom, where the red dashed and green dot-dashed lines denote the numerical simulations of Eq. (\ref{1-4}) and analytical results based
on Eq. (\ref{1-5}), respectively. We find that atomic excitation dynamics of bound states with two frequencies exhibits persistent oscillation.

When $d=\left(  2n+1\right)  \frac{\pi}{\Omega},$ $n\in N\Rightarrow d\sim
10^{2}\lambda$, Fig. \ref{fig4}(a), demonstrates symmetric pole distribution around frequency $\alpha$, resulting in a periodic oscillation with uniform amplitude. Complex poles significantly expedite the attainment of a steady state. A larger separation between coulping points corresponds to a higher density of complex poles, facilitating a quicker convergence to a steady state with reduced amplitude of stable fluctuations.
The atomic excitation probability $\left\vert C_{e}\left(  t\right)  \right\vert ^{2}$ varies with $\gamma\tau$. Notably, despite changes in the coupling points' separation, the probability remains at a corresponding constant amplitude. Intriguingly, quantum interference between these points induces periodic oscillations in the dynamics. These oscillations persist with a frequency of $2\Omega$, showing no decay in amplitude over time, show in Fig. \ref{fig4}(d).
In the detailed dressed state depiction, the giant three-level atom exhibits two unique dressed states, establishing a stable bound state at each level. This sets the groundwork for interference between the bound states, giving rise to periodic interference patterns in the dynamic evolution of giant atoms.

\section{CONCLUSION AND DISCUSSION}
In this paper, we investigate the coupling between a giant $\Lambda $-type three-level atom and a one-dimensional waveguide. We analyze the evolution of spontaneous emission dynamics of the giant atomic excited state from both analytical and numerical perspectives.
We have shown that the distance between coupling points can be longer than the characteristic wavelength of the
bath, i.e., $d=\frac{2n+1}{\alpha\pm\Omega}\pi\sim\frac{2\pi}{\alpha}$ for $n$ is a smaller positive integer. It is necessary to consider the phase difference between
these coupling points. The discrete distance between
two coupling points, a non-Markovian effect can be observed, spontaneous
decay is polynomial instead of exponential owing to the appearance of
bound state. More specifically, we have demonstrated that the energy level configuration of a giant atom must meet specific constraints to support two oscillatory bound states. In this article, we simplify the discussion by assuming $\phi=0$ , $\Delta=0$ and $\alpha/\Omega=q, q\in N$ without loss of generality.

When $q$ is an odd integer, a series of discrete distances $d=\frac{2n+1}{\alpha\pm\Omega}\pi$ can ensure the satisfaction of a bound state while also encompassing a quasi-bound state. The interference between these two states induces oscillations in the atom's spontaneous emission. As the distance increases, the quasi-bound state decays rapidly, leading to the disappearance of interference and oscillations. Essentially, the atomic probability tends to stabilize within a shorter time frame. Furthermore, the magnitude of this stable value is distance-dependent. Greater distances result in a decreased probability of the atom being confined in the excited state.

When $q$ is an even integer, a series of discrete distances $d=\frac{2n+1}{\Omega}\pi$ can accommodate two bound states simultaneously. In this context, the number of quasi-bound states varies with distance and lacks uniqueness. Both bound and quasi-bound states exhibit symmetry relative to the frequency $\Omega$. Unlike the scenario with a single bound state, in this case, the distances are considerably large, leading to minimal impact from the quasi-bound states which fade quickly. The subsequent evolution of the atomic excited state probability is solely governed by the two bound states. A consistent, equal-amplitude periodic oscillation with a period of $\Omega$, independent of distance, occurs between the two bound states. While theoretically all ranges of distances can manifest the described scenario, in practical physics, the atom's size significantly exceeds the wavelength of the waveguide, often by over 100-fold, necessitating higher precision experimental instruments to detect the atom's probability approaching zero. Therefore, when truncating to $n=2$, the atom tends to decay rapidly to zero, hindering oscillations. Essentially, in scenarios with distant coupling points, the atom can be viewed as a small atom coupled at a single point in an infinitely long waveguide. Over an extended period, the atom's spontaneous emission drives the probability of the atom being in an excited state towards zero.

Through the analysis of interference phenomena caused by the giant atom, we observe that the overall decay rate of spontaneous emission and the final stable value change with the size of the giant atom. The larger the atomic size, the faster the decay, leading to a lower final stable value or amplitude of stable oscillations. Finally, the oscillating bound state, i.e., the dynamical exchange of excitations between the atom excited state $\left\vert e\right\rangle $
 and the metastable state $\left\vert s\right\rangle $,
essentially demonstrates the Rabi-oscillation phenomenon of cavity QED. Typically, cavity QED requires two mirrors, but this shows that a single $\Lambda $-type three-level atom with two coupling points in the open waveguide is a minimalistic implementation of cavity QED.

\begin{acknowledgments}
This work was supported by NSFC Grants No.11935006, No. 12075082, No. 12247105,
the science and technology innovation Program of Hunan Province (Grant No. 2020RC4047),
XJ2302001, and the Science and Technology Department of Hunan Provincial Program (2023ZJ1010).
\end{acknowledgments}

\section*{APPENDIX A: THE EQUATIONS OF MOTION IN THE SINGLE EXCITATION
SUBSPACE}

In this appendix, we show the details of how to obtain the equation of
motion for the probability amplitude of the giant atom being excited. From
the Schr\"{o}inger equation $\mathrm{i}\frac{\partial }{\partial t}%
\left\vert \Psi \left( t\right) \right\rangle =H\left\vert \Psi \left(
t\right) \right\rangle $, we have \setcounter{equation}{0} %
\renewcommand{\theequation}{A\arabic{equation}}

\begin{widetext}
\begin{subequations}
\label{A-1}
\begin{align}
\frac{d}{dt}C_{e}\left(  t\right)    & =-\mathrm{i}\alpha_{e}C_{e}\left(
t\right)  -\mathrm{i}\Omega C_{s}\left(  t\right) -\mathrm{i}\int_{-\infty}^{+\infty}dkg_{0}\sqrt{\left\vert k\right\vert
}\left(  1+e^{\mathrm{i}kd}\right)  \beta_{k}\left(  t\right)  ,\label{A-1a}\\
\frac{d}{dt}C_{s}\left(  t\right)    & =-\mathrm{i}\alpha_{s}C_{s}\left(
t\right)  -\mathrm{i}\Omega C_{e}\left(  t\right)  ,\\
\frac{d}{dt}\beta_{k}\left(  t\right)    & =-\mathrm{i}\left(  \omega
_{g}+\tilde{\omega}_{k}\right)  -\mathrm{i}g_{0}\sqrt{\left\vert k\right\vert
}\left(  1+e^{-\mathrm{i}kd}\right)  C_{e}\left(  t\right)  ,
\end{align}
\end{subequations}
Integrating Eq. (\ref{A-1}b) and Eq. (\ref{A-1}c), we have the formal
solutions
\begin{subequations}
\label{A-2}%
\begin{align}
C_{s}\left(  t\right)    & =C_{s}\left(  0\right)  e^{-\mathrm{i}\alpha_{s}%
t}-\mathrm{i}\Omega\int_{0}^{t}dt^{\prime}C_{e}\left(  t^{\prime}\right)
e^{\mathrm{i}\alpha_{s}\left(  t^{\prime}-t\right)  },\label{A-2a}\\
\beta_{k}\left(  t\right)    & =\beta_{k}\left(  0\right)  e^{-\mathrm{i}%
\left(  \omega_{g}+\tilde{\omega}_{k}\right)  t} -\mathrm{i}g_{0}\sqrt{\left\vert k\right\vert }\left(  1+e^{-\mathrm{i}%
kd}\right)  \int_{0}^{t}dt^{\prime}C_{e}\left(  t^{\prime}\right)
e^{\mathrm{i}\left(  \omega_{g}+\tilde{\omega}_{k}\right)  \left(  t^{\prime
}-t\right)  }.
\end{align}
\end{subequations}
Inserting Eq. (\ref{A-2}a)-(\ref{A-2}b) into Eq. (\ref{A-1}a), and we assume that the giant atom (waveguide) is initially in the excited
$\left\vert e\right\rangle $ (vacuum $\left\vert 0\right\rangle $) state,
i.e., $\left\vert \psi\left(  0\right)  \right\rangle =\left\vert
e,0\right\rangle .$ Then, we obtain the following differential
equation:

\begin{align}
\frac{d}{dt}C_{e}\left(  t\right)    & =-\mathrm{i}\alpha_{e}C_{e}\left(
t\right)  -\Omega^{2}\int_{0}^{t}dt^{\prime}C_{e}\left(  t^{\prime}\right)
e^{\mathrm{i}\alpha_{s}\left(  t^{\prime}-t\right)  }\nonumber\\
& -g_{0}^{2}\int_{-\infty}^{+\infty}dk\left\vert k\right\vert \left(
2+e^{\mathrm{i}kd}+e^{-\mathrm{i}kd}\right)  \int_{0}^{t}dt^{\prime}%
C_{e}\left(  t^{\prime}\right)  e^{\mathrm{i}\left(  \omega_{g}+\tilde{\omega
}_{k}\right)  \left(  t^{\prime}-t\right)  }.\label{A-3}%
\end{align}

In order to simplify Eq. (\ref{A-3}), we adopt Weisskopf-Wigner approximation, that is to say,
under the
rotating-wave approximation, the processes we focuses on in involve a narrow range of wave vectors around the atomic transition frequency $\omega
_{eg}=\omega _{e}-\omega _{g}$. Therefore, the boundary of the integral can be changed to $\int_{-\infty }^{+\infty
}$, and $\omega_{k}$ can be
replaced by the constant $\omega _{eg}$, which is mentioned outside the
Integral symbol. We define the relaxation rate
\begin{equation}
\gamma\equiv4\pi\frac{g_{0}^{2}\omega_{eg}}{v^{2}},
\label{A-4}
\end{equation}
and we have defined the delay time term
\begin{equation}
\frac{g_{0}^{2}}{v^{2}}\int_{-\infty}^{+\infty}\omega_{k}e^{\mathrm{i}%
\omega_{k}\left[  d/v+\left(  t^{\prime}-t\right)  \right]  }d\omega
_{k}\approx\frac{2\pi g_{0}^{2}\omega_{eg}}{v^{2}}\delta\left(  t^{\prime
}-t+\tau\right)  \rightarrow\frac{\gamma}{2}\delta\left(  t^{\prime}%
-t+\tau\right)  .
\label{A-5}
\end{equation}
Therefore, the dynamics for the giant atom is
\begin{equation}
\frac{d}{dt}C_{e}\left( t\right) =-\mathrm{i}\alpha _{e}C_{e}\left( t\right)
-\Omega ^{2}\int_{0}^{t}dt^{\prime }\beta \left( t^{\prime }\right) e^{-%
\mathrm{i}\omega _{s}\left( t-t^{\prime }\right) }-\gamma C_{e}\left(
t\right) -\gamma C_{e}\left( t-\tau \right) e^{-\mathrm{i}\phi }\Theta
\left( t-\tau \right) .  \label{A-6}
\end{equation}
In order to solve Eq. (\ref{A-6}), we use Laplace transformation $\mathscr{L}%
\left[ C_{e}\left( t\right) \right] =\tilde{C}_{e}\left( s\right)
=\int_{0}^{+\infty }dtC_{e}\left( t\right) e^{-st}$ and obtain%
\begin{equation}
s\tilde{C}_{e}\left( s\right) -C_{e}\left( 0\right) =\left( -\mathrm{i}%
\alpha _{e}-\gamma \right) \tilde{C}_{e}\left( s\right) -\gamma e^{-\mathrm{i%
}\phi }e^{-s\tau }\tilde{C}_{e}\left( s\right) -\Omega ^{2}\tilde{C}%
_{e}\left( s\right) \frac{1}{s+\mathrm{i}\alpha _{s}},  \label{A-7}
\end{equation}
in other words
\begin{equation}
\tilde{C}_{e}\left( s\right) =\left[ s+\mathrm{i}\alpha _{e}+\gamma +\gamma
e^{-\mathrm{i}\phi }e^{-s\tau }+\frac{\Omega ^{2}}{s+\mathrm{i}\omega _{s}}%
\right] ^{-1}.  \label{A-8}
\end{equation}

So, the time evolution of $C_{e}\left( t\right) $ can be obtained by the
inverse Laplace transformation, the key lies in the poles of $\tilde{C}%
_{e}\left( s\right) ,$ it given by the roots of the following equation:
\begin{equation}
f\left( s_{p}\right) =s_{p}+\mathrm{i}\alpha _{e}+\gamma +\gamma e^{-\mathrm{%
i}\phi }e^{-s_{p}\tau }+\frac{\Omega ^{2}}{s_{p}+\mathrm{i}\omega _{s}}=0.
\label{A-9}
\end{equation}

The derivative function of the function on the left side of the Eq. (\ref%
{A-9}) for $s$ is
\begin{equation}
f^{\prime }\left( s\right) =1-\gamma \tau e^{-\mathrm{i}\phi }e^{-s\tau }-%
\frac{\Omega ^{2}}{\left( s+\mathrm{i}\alpha _{s}\right) ^{2}},  \label{A-10}
\end{equation}%
the explicit form of $C_{e}\left( t\right) $ is thus%
\begin{equation}
C_{e}\left( t\right) =\sum\limits_{p}\frac{e^{s_{p}t}}{f^{\prime }\left(
s_{p}\right) }.  \label{A-11}
\end{equation}
and the time evolution of the bosonic field function $\varphi \left(
x,t\right) \equiv \frac{1}{\sqrt{2\pi }}\int_{-\infty }^{+\infty }dke^{%
\mathrm{i}kx}\beta _{k}\left( t\right) $ in the waveguide is
\begin{align}
\varphi\left(  x,t\right)    & =-\mathrm{i}g_{0}\frac{1}{\sqrt{2\pi}}%
\int_{-\infty}^{+\infty}dke^{\mathrm{i}kx}\sqrt{\left\vert k\right\vert
}\left(  1+e^{-\mathrm{i}kd}\right)  \int_{0}^{t}dt^{\prime}C_{e}\left(
t^{\prime}\right)  e^{\mathrm{i}\left(  \omega_{g}+\tilde{\omega}_{k}\right)
\left(  t^{\prime}-t\right)  }\nonumber\\
& \approx-\mathrm{i}\sqrt{\frac{\gamma}{2}}\left[  C_{e}\left(  t-\frac
{\left\vert x-d\right\vert }{v}\right)  e^{-\mathrm{i}\phi\frac{\left\vert
x-d\right\vert }{d}}\Theta\left(  t-\frac{\left\vert x-d\right\vert }%
{v}\right)  +C_{e}\left(  t-\frac{\left\vert x\right\vert }{v}\right)
e^{-\mathrm{i}\phi\frac{\left\vert x\right\vert }{d}}\Theta\left(
t-\frac{\left\vert x\right\vert }{v}\right)  \right]  ,\label{A-12}%
\end{align}
\end{widetext}
where $\phi=\left(  \omega_{g}-\frac{\omega_{l}}{2}\right)  \tau$. Besides the atomic population on the metastable
state $\left\vert s\right\rangle $ can be then written as
\begin{equation}
C_{s}\left( t\right) =-\mathrm{i}\Omega \int_{0}^{t}dt^{\prime }C_{e}\left(
t^{\prime }\right) e^{\mathrm{i}\alpha _{s}\left( t^{\prime }-t\right) }.
\label{A-13}
\end{equation}
which are actually Eqs. (\ref{A-11})-(\ref{A-13}) in the main text.

\section*{APPENDIX B: DARK-STATE CONDITION}
\setcounter{equation}{0}

\renewcommand{\theequation}{B\arabic{equation}}

To obtain the explicit solutions for $C_{s}\left( t\right) $ and $\varphi\left(  x,t\right)$, we first need to solve Eq. (\ref{A-11}). The key to solve Eq. (\ref{A-11}) is to find the poles of $f\left( s_{p}\right)=0$, i.e., find the roots of the transcendental equation
\begin{equation}
s_{p}+\mathrm{i}\alpha _{e}+\gamma +\gamma e^{-\mathrm{i}\phi }e^{-s_{p}\tau
}+\frac{\Omega ^{2}}{s_{p}+\mathrm{i}\omega _{s}}=0.  \label{B-1}
\end{equation}
Physically, the complex frequency $s_{p}$ has a negative real part, which represents the decay rate. And $s_{p}$ has a negative imaginary part, which represents the self-energy of
the atom. If $s_{p}$ is purely imaginary, the corresponding mode is a dark state, which does not decay. We seek the purely imaginary solution $s_{p}\equiv -\mathrm{i}\Omega _{p}$, then we have the equations
\begin{subequations}
\label{B-2}%
\begin{align}
0  & =\gamma+\gamma\cos\left(  \Omega_{p}\tau-\phi\right)  ,\label{B-2a}\\
0  & =\gamma\sin\left(  \Omega_{p}\tau-\phi\right)  -\left(  \Omega_{p}%
-\alpha_{e}\right)  +\frac{\Omega^{2}}{\Omega_{p}-\alpha_{s}}.
\end{align}
\end{subequations}
The constraint shown in Eq. (\ref{B-2a}) has a series of solutions yields $%
\cos \left( \Omega _{p}\tau -\phi \right) =-1\Leftrightarrow \Omega _{p}\tau
=\left( 2p+1\right) \pi +\phi ,p\in Z$. And with it comes that
\begin{equation}
\left( \Omega _{p}-\alpha _{e}\right) -\frac{\Omega ^{2}}{\Omega _{p}-\alpha
_{s}}=0.  \label{B-3}
\end{equation}

Getting
\begin{subequations}
\label{B-6}
\begin{eqnarray}
\Omega _{p_{1}} &=&\frac{\left( \alpha _{e}+\alpha _{s}\right) +\sqrt{\left(
\alpha _{e}-\alpha _{s}\right) ^{2}+4\Omega ^{2}}}{2}, \\
\Omega _{p_{2}} &=&\frac{\left( \alpha _{e}+\alpha _{s}\right) -\sqrt{\left(
\alpha _{e}-\alpha _{s}\right) ^{2}+4\Omega ^{2}}}{2}.
\end{eqnarray}%
It is inevitable to find two integers $p_{1}$ and $p_{2}$\ when $\Omega \neq
0$. This means that, the atomic excitation probability amplitude $%
C_{e}\left( t\right) $ can be written as
\end{subequations}
\begin{subequations}
\label{B-7}
\begin{eqnarray}
f^{\prime }\left( -\mathrm{i}\Omega _{p_{1}}\right) &=&\gamma \tau +\frac{1}{%
\cos ^{2}\theta }, \\
f^{\prime }\left( -\mathrm{i}\Omega _{p_{2}}\right) &=&\gamma \tau +\frac{1}{%
\sin ^{2}\theta }.
\end{eqnarray}
\end{subequations}

Here, we can consider that the coupling between the transition $\left\vert
e\right\rangle \leftrightarrow\left\vert s\right\rangle $ and the classical
driving field leads to dressed states in single exciton space. The expression
of the dressed states can be obtained as
\begin{subequations}
\label{B-4}
\begin{align}
\left\vert +\right\rangle  & =\cos\theta\left\vert e\right\rangle +\sin
\theta\left\vert s\right\rangle ,\label{B-4a}\\
\left\vert -\right\rangle  & =-\sin\theta\left\vert e\right\rangle +\cos
\theta\left\vert s\right\rangle ,
\end{align}
\end{subequations}
with the detuning $\Delta\equiv$ $\alpha_{e}-\alpha_{s},$ and the effect
frequency $\Omega_{eff}\equiv\sqrt{\left(  \alpha_{e}-\alpha_{s}\right)
^{2}+4\Omega^{2}}.$ Besides, the eigenvalues are
\begin{equation}
\omega_{\pm}=\left(  \Delta\pm\Omega_{eff}\right)  /2,
\label{B-5}
\end{equation}
and the corresponding stacking weight is
\begin{subequations}
\label{B-6}%
\begin{align}
\sin\theta & =\sqrt{\frac{\Omega_{eff}-\Delta}{2\Omega_{eff}}},\label{B-6a}\\
\cos\theta & =\sqrt{\frac{\Omega_{eff}+\Delta}{2\Omega_{eff}}}.
\end{align}
\end{subequations}
Thus, if two dark state solutions coexist, then the expression for $C_{e}\left(  t\right)  $ is
\begin{equation}
C_{e}\left( t\right) =\frac{\cos ^{2}\theta }{\gamma \tau \cos ^{2}\theta +1}%
e^{-\mathrm{i}\Omega _{p_{1}}t}+\frac{\sin ^{2}\theta }{\gamma \tau \sin
^{2}\theta +1}e^{-\mathrm{i}\Omega _{p_{2}}t}.  \label{B-8}
\end{equation}

To simplify the discussion, and without losing generality, let's assume
$\Delta=0\Leftrightarrow\alpha_{e}=\alpha_{s}=\alpha\Rightarrow$ $\sin
\theta=\cos\theta=\sqrt{1/2},$ and $\omega_{g}=\omega_{l}/2\Rightarrow\phi=0,$
so the two dark modes are%
\begin{subequations}
\label{B-9}%
\begin{align}
\Omega_{p_{1}}  & =\alpha+\Omega=\left(  2p_{1}+1\right)  \pi/\tau
,\label{B-9a}\\
\Omega_{p_{2}}  & =\alpha-\Omega=\left(  2p_{2}+1\right)  \pi/\tau.
\end{align}
\end{subequations}
Where $p_{1},p_{2}\in N$. If one of the conditons $\left\{  \Omega_{p_{1}},\Omega_{p_{2}}\right\}  $ is
satisfied, and there is only one integer $p_{1}$ or $p_{2}$ as its solution.
In this case, the excited probability amplitude of the giant atom is
\begin{subequations}
\label{B-10}%
\begin{align}
C_{e}\left(  t\right)    & =\frac{1}{\gamma\tau+2}e^{-\mathrm{i}\Omega_{p_{1}%
}t},\label{B-10a}\\
C_{e}\left(  t\right)    & =\frac{1}{\gamma\tau+2}e^{-\mathrm{i}\Omega_{p_{2}%
}t}.
\end{align}
\end{subequations}

And inserting the dark-states solution $C_{e}\left(  t\right)  $ into Eq.
(\ref{A-13}), we obtain the long-time dynamics of the atomic probability
amplitude in the metastable state
\begin{subequations}
\label{B-10}%
\begin{align}
C_{s}\left(  t\right)   &  =\frac{1}{\gamma\tau+2}\left(  e^{-\mathrm{i}%
\Omega_{p_{1}}t}-e^{-\mathrm{i}\alpha t}\right)  ,\label{B-11a}\\
C_{s}\left(  t\right)   &  =-\frac{1}{\gamma\tau+2}\left(  e^{-\mathrm{i}%
\Omega_{p_{2}}t}-e^{-\mathrm{i}\alpha t}\right)  .
\end{align}
\end{subequations}

Inserting Eqs. (\ref{B-10a})-(\ref{B-10}b) into Eq.
(\ref{A-12}), we obtain the expression for the field density in the long-time limit
\begin{subequations}
\label{B-12}%
\begin{align}
\varphi\left(  x,t\right)   &  =-\mathrm{i}\sqrt{\frac{\gamma}{2}}\frac
{1}{\gamma\tau+2}\left[  e^{-\mathrm{i}\Omega_{p_{1}}\left(  t-\frac
{\left\vert x-d\right\vert }{v}\right)  }+e^{-\mathrm{i}\Omega_{p_{1}}\left(
t-\frac{\left\vert x\right\vert }{v}\right)  }\right]  ,\label{B-12a}\\
\varphi\left(  x,t\right)   &  =-\mathrm{i}\sqrt{\frac{\gamma}{2}}\frac
{1}{\gamma\tau+2}\left[  e^{-\mathrm{i}\Omega_{p_{2}}\left(  t-\frac
{\left\vert x-d\right\vert }{v}\right)  }+e^{-\mathrm{i}\Omega_{p_{2}}\left(
t-\frac{\left\vert x\right\vert }{v}\right)  }\right]  .
\end{align}
\end{subequations}
If two integers $p_{1}$ and $p_{2}$ satisfied $p_{1},p_{2}\in N$
simultaneously. Eqs. (\ref{B-9a})-(\ref{B-9}b) gives an additional condition
for the coexisting dark states, i.e.,%
\begin{equation}
\tau=\frac{\left(  2p_{1}+1\right)  \pi}{\alpha+\Omega}=\frac{\left(
2p_{2}+1\right)  \pi}{\alpha-\Omega}, \label{B-13}%
\end{equation}
In the energy level configuration of a three-level atom, we assumed
$q\equiv\alpha/\Omega$ $\left(  q\gg1\right)  $, as given by the rotating-wave
approximation. Combining this with Eq. (\ref{B-13}), we find that the
solutions are of the form
\begin{subequations}
\label{B-14}%
\begin{align}
p_{1}  &  =\frac{q}{2}+\left(  q+1\right)  n,\label{B-14a}\\
p_{2}  &  =\frac{q}{2}-1+\left(  q-1\right)  n,
\end{align}
\end{subequations}
with $q\in\left\{  q|q=2x,x\in N^{\ast}\right\}  $ and $n\in N$. The
conditions in Eqs. (\ref{B-14a})-(\ref{B-14}b) then become
\begin{equation}
d=\left(  2n+1\right)  \frac{\pi}{\Omega},\label{B-15}%
\end{equation}
It is worth noting that if $q$ is an odd integer, the constraint condition
shown by Eq. (\ref{B-13}) cannot be established,  then two bound states cannot
coexist. Of course, if $q$ is a decimal, corresponding constraints can also be
found. We don't discuss it here.

%%%%%%%%%%%%%%%%%%%%%%%%%%%%%%%%%%%%%%%%%%%%%%%%%%%%%%%%%%%%

\end{document}